\documentclass[english,onecolumn,prx,preprintnumbers,amsmath,amssymb,superscriptaddress]{revtex4-2}
\usepackage{verbatim}
\usepackage{graphicx}
\usepackage{amssymb}
\usepackage{ulem}
\usepackage{xcolor}

\usepackage{babel}
\usepackage{float}
\usepackage{titlesec}
\makeatletter

\def\maketitle{
	\@author@finish
	\title@column\titleblock@produce
	\suppressfloats[t]}
\makeatother

\makeatother

\begin{document}
\preprint{XXX}

\newcommand{\Vac}{V_{\mathrm{ac}}}
\newcommand{\Vdc}{V_{\mathrm{dc}}}
\newcommand{\omd}{\omega_d}

\title{Tunable Fano Resonance and Frequency Locking in a Graphene-SiNx Hybrid Nanomechanical Resonator}

\author{Ateesh K. Rathi}
\affiliation{Department of Physics, Indian Institute of Technology - Kanpur, UP-208016, India}

\author{Javed A. Mondal}
\affiliation{Department of Physics, Indian Institute of Technology - Kanpur, UP-208016, India}

\author{Rajan Singh}
\affiliation{Department of Physics, Indian Institute of Technology - Kanpur, UP-208016, India}

\author{Ryan J.T. Nicholl}
\affiliation{Department of Physics and Astronomy, Vanderbilt University, Nashville, Tennessee 37235, USA}	

\author{Kirill I. Bolotin}
\affiliation{Department of Physics, Freie Universitat Berlin, Arnimallee 14, Berlin 14195, Germany}

\author{Saikat Ghosh}
\email{gsaikat@iitk.ac.in}
\affiliation{Department of Physics, Indian Institute of Technology - Kanpur, UP-208016, India}

%\author{Sagar Chakraborty}

%\keywords{graphene, nanoelectromechanical systems, internal resonance, phononic frequency comb, high-harmonic generation, period doubling}

\date{\today}

\begin{abstract}
Fano resonances, arising from the interference between discrete and continuum
states, are observed across a wide range of quantum and classical systems.
Here, we report the experimental observation of Fano resonances in a
graphene SiNx hybrid nanomechanical system modeled as coupled oscillators.
The broad, low quality factor graphene mode plays the role of the continuum,
while the dense comb of sharp, high quality factor SiNx modes provides the
discrete states. The inter-mode detuning is tunable via a DC gate voltage,
enabling dynamic control of the Fano resonance: we demonstrate gate controlled
switching of both the sign and the magnitude of the Fano asymmetry parameter
$q$, in quantitative agreement with a coupled oscillator theory that predicts
$q=-\cot\phi$, with $\phi$ the phase of the continuum response. At strong
drive, the graphene mode enters the Duffing regime and its jump-down
frequency locks to successive SiNx modes, producing a staircase of
drive insensitive frequency plateaus; a weak seeding tone deterministically
switches the resonator between adjacent locked states. The dense SiNx
mode thus acts, in the linear regime, as the discrete states of a
tunable Fano interferometer and, in the nonlinear regime, as a frequency
ruler that stabilizes and quantizes the graphene oscillation. This platform
offers a controllable mechanical realization of Fano interference and opens
new avenues for high resolution hybrid resonant sensors and stable
nanomechanical frequency references.
\end{abstract}
\maketitle

\section*{Introduction}
Fano resonance results from the interference between a discrete quantum state
and a continuum of states~\cite{Fano1935,PhysRev.124.1866}. Although originally
regarded as a purely quantum effect, it is now understood as a universal
wave interference phenomenon with observable manifestations in classical
systems~\cite{Rodrigues2009,Stassi2017,Chen2021,Xie2025}, including
optics~\cite{Fan2003}, acoustics, and metamaterials. The distinctive
asymmetric line shape associated with the Fano resonance has enabled diverse
optical applications, including sensors~\cite{Limonov:21,ElKabbash2021},
optical filters and switches~\cite{Rybin2015}, broadband
reflectors~\cite{Chang2013}, lasers~\cite{Yu2016}, and slow-light
devices~\cite{PhysRevE.74.046603,Chang2013}.  In classical systems, Fano resonances have been investigated for both fundamental physics and sensing applications~\cite {Stassi2019}. Nearly two decades ago, theoretical studies predicted that Fano resonances could emerge in two weakly coupled mechanical harmonic oscillators subjected to periodic driving~\cite{Joe2006}. These mechanical analogs are now commonly used to illustrate Fano interference in quantum~\cite{RevModPhys.82.2257} and optical~\cite{Limonov2017} systems. This perspective also motivates studies of nanomechanical resonators driven beyond their linear regime.

When operated beyond the linear response regime, nanomechanical resonators display significant nonlinear dynamics. In graphene resonators, geometric nonlinearities lead to Duffing type behavior at comparatively low drive amplitudes~\cite{Chen2009,Eichler2011}. In this regime, the resonance frequency depends on amplitude, so amplitude fluctuations are converted into frequency fluctuations. This process reduces the frequency stability and sensing performance of the resonator.
%A complementary body of work concerns nanomechanical resonators driven beyond their linear regime. Atomically thin membranes such as graphene enter the Duffing (hardening) regime at very modest drive levels~\cite{Chen2009,Eichler2011}, where the amplitude frequency (a--f) effect converts amplitude fluctuations into frequency noise, degrading their performance as oscillators and sensors. 
In this context, locking a nonlinear resonator to an auxiliary stable mode via internal resonance~\cite{Antonio2012}, synchronization~\cite{Shim2007,PhysRevLett.112.014101}, or external injection locking~\cite{Houri2017} has been shown to suppress this noise conversion and stabilize the oscillation frequency.

Here, we report the experimental observation of Fano resonance in a weakly
coupled graphene-SiNx hybrid nanomechanical resonator, together with a
locking phenomenon that emerges when the same device is driven into the
nonlinear regime. In the linear regime, the interference profile is shown to
be effectively tunable by varying the resonator stiffness through DC and AC
gate voltages. A coupled oscillator model, solved analytically and numerically verified
 in the Supplementary Material (SM)~\cite{supplementFano}, reproduces all
observed signatures: asymmetric line shapes, gate controlled sign
reversal, and the relation between the Fano parameter and the interference
phase. At strong drive, the jump-down frequency of the graphene Duffing
response locks to the fixed SiNx modes, tracing a staircase of
drive insensitive plateaus that can be stepped deterministically with a weak
seeding tone. These findings provide direct evidence of tunable destructive
interference in a mechanical platform and suggest new opportunities for
high performance graphene based resonant sensors and frequency stabilized
nanomechanical oscillators.

\section*{Experimental setup}
\begin{figure*}[tb]
\includegraphics[width=\textwidth]{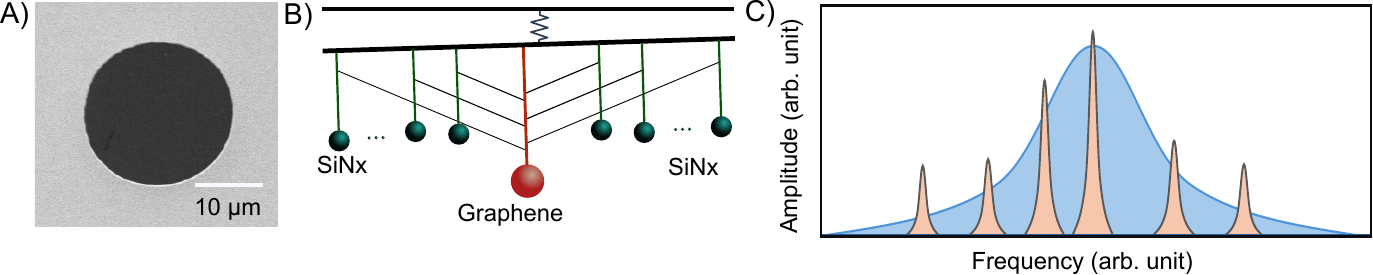}
\caption{\textbf{(A)}~SEM image of the device, where the grey area corresponds to the silicon nitride membrane, while the black circular region represents a suspended graphene drum with a diameter of 20~$\mu$m. \textbf{(B)}~Mechanical model of the system: a single damped graphene resonator (red sphere) coupled to multiple high-$Q$ silicon nitride resonators (teal spheres). \textbf{(C)}~Schematic of the broad graphene resonance overlapping with a dense comb of sharp, fixed frequency silicon nitride modes.}
\label{fig:model}
\end{figure*}

\begin{figure*}[tb]
\includegraphics[width=\textwidth]{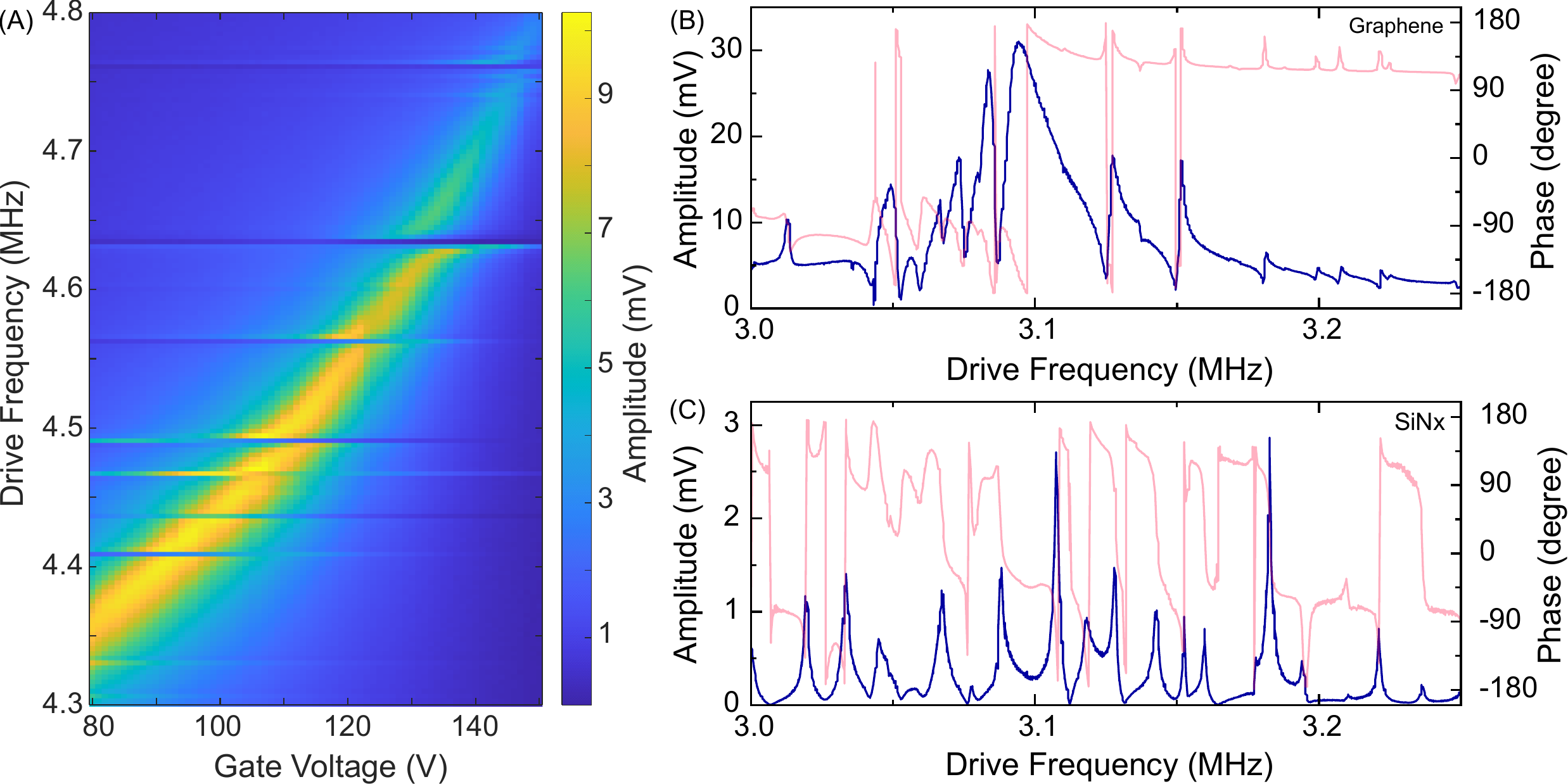}
\caption{\textbf{(A)}~Dispersion of the graphene resonator resonance frequency as a
function of DC gate voltage $V_{dc}$. The tunable graphene mode successively
interacts with multiple fixedfrequency silicon nitride (SiNx) modes (dark
horizontal lines), which remain unaffected by the gate voltage due to the
large mass and high built-in tension of SiNx. \textbf{(B)}~Representative frequency domain spectrum of the graphene resonator at a
fixed gate voltage, showing the broad graphene resonance strongly coupled to
multiple discrete, high quality factor silicon nitride modes. \textbf{(C)}~Spectrum in the same frequency range, highlighting the dense population of sharp silicon
nitride modes available for coupling with the broad graphene resonance,
enabling near one-to-one mode interactions across the graphene linewidth. Blue and red curves in both (B) and (C) represent amplitude and phase response respectively, displaying clear phase jumps associated with nearly every silicon nitride mode.}
\label{fig:fig2}
\end{figure*}

The experimental realization of the tunable Fano resonance was achieved in a
graphene--SiNx hybrid resonator. The hybrid system consists of a
large area ($320 \times 320 \times 0.3~\mu\mathrm{m}^3$) SiNx membrane
coated with a $20$~nm thick gold (Au) layer, which contains holes with
diameters of $20$~$\mu$m. Monolayer graphene is mechanically transferred
onto the entire structure. Fig.~\ref{fig:model}(A) shows an SEM image of the device. A conductive silicon chip is positioned
below the graphene SiNx assembly, separated by an insulating layer.
Separate electrical contacts are established for the SiNx/Au/graphene
stack and for the bottom silicon, and the complete structure is placed within
a vacuum chamber. The mechanical motion of the suspended graphene and of the
SiNx membrane is independently actuated and detected. To probe the motion,
a fiber based confocal microscope is focused on either the graphene or the
SiNx; the motion induced modulation of the reflected signal is amplified
and recorded using a Michelson interferometer. Graphene is actuated electrostatically by applying an oscillating voltage between the membrane and the gating chip. At the power levels used in this study, this modulation is insufficient to directly actuate the significantly heavier SiNx membrane. Experiments were conducted using two distinct samples. The data shown in Figures~\ref{fig:fig2}(B) and~\ref{fig:fig2}(C) were collected from one sample, while all other experimental results were obtained from another sample of the same type.

A parallel plate capacitor is formed between the graphene SiNx layer and
the conductive silicon chip. The hybrid resonator is biased by a DC gate
voltage $V_{dc}$ across the plates, and its stiffness is tuned by this gate
voltage. Both the DC bias $V_{dc}$ and an AC voltage of amplitude
$V_{ac}\cos(\omega t)$ are applied across the capacitor, producing the driving
force
\begin{equation}
F \approx \frac{1}{2}\,\frac{\epsilon_0 A}{d^{2}}
\left(V_{dc}+V_{ac} \cos(\omega t)\right)^{2},
\label{eq:force}
\end{equation}
which expands to
\begin{equation}
F=\frac{1}{2}\,\frac{\epsilon_0 A}{d^{2}}
\left(V_{dc}^{2}+2V_{ac}V_{dc}\cos(\omega t)+V_{ac}^{2}\cos^{2}(\omega t)\right).
\label{eq:force2}
\end{equation}
Here, $\epsilon_0$ denotes the vacuum permittivity, $A$ the overlap area, and
$d$ the spacing between the driving electrode and the graphene SiNx
membrane. In Eq.~(\ref{eq:force2}), the first term is a static force that
shifts the resonance frequency of the graphene resonator, as illustrated in
Fig.~\ref{fig:fig2}(A); at this power level the SiNx membrane
remains stationary. The last term produces components at $2\omega$ and DC,
which can be neglected when $V_{ac}<< V_{dc}$, while the second term represents the required AC
driving force. Consequently, the time varying force applied to the graphene
resonator is
\begin{equation}
F \approx \frac{\epsilon_0 A}{d^{2}}\,V_{ac}V_{dc}\cos(\omega t)
= F_0\cos(\omega t),
\label{eq:drive}
\end{equation}
with $F_0=\epsilon_0 A V_{ac} V_{dc}/d^{2}$, demonstrating that a harmonic force
is produced.

The micromechanical hybrid resonator was mounted in a clamp for electrical
connection and placed in a custom vacuum chamber held at~mTorr. A linear
DC power supply provided the DC bias for the resonators, and a lock-in
amplifier (Zurich Instruments HF2LI) was used to the frequency sweep signal and
measured the output.

The large lateral dimensions and high tensile stress of the silicon nitride membrane enable a wide range of vibrational modes, with quality factors considerably higher than those of graphene resonators.
Within this system, the graphene resonator's significant damping results in a broad frequency response, effectively behaving as a continuum of states in the context of Fano interference, whereas silicon nitride, owing to its high quality factor, exhibits sharp,
well defined resonance peaks. Fig.~\ref{fig:model}(B) presents a schematic
model of the hybrid system: the large red sphere represents the graphene resonator, while the teal spheres denote individual silicon nitride resonators. To account for the
high mode density of SiNx, multiple resonators are included, each coupled
to the graphene resonator with a distinct coupling strength (shown by the black lines). The SiNx modes are not directly coupled to each other. Slight changes in the length of each SiNx resonator represent different natural resonance frequencies. Fig.~\ref{fig:fig2}(A) illustrates that many silicon nitride modes
lie within the frequency range of the broad graphene resonance: some SiNx
modes are located near the center of the graphene resonance, while others
extend into the tails of its profile, as sketched in
Fig.~\ref{fig:model}(C).

The resonance frequency of the graphene resonator can be tuned by adjusting
the DC gate voltage $V_{dc}$. In contrast, the effect of this gate voltage on
the silicon nitride resonators is negligible because of their much larger
mass and built-in tension; the SiNx mode frequencies therefore remain
unaffected. In Fig.~\ref{fig:fig2}(A), the gate voltage is swept from 80~V
to 150~V, shifting the graphene resonance across this range and producing
successive interactions with the different fixed frequency silicon nitride
modes (dark horizontal lines).
At a fixed gate voltage, the spectrum of the graphene resonator reveals
interactions with multiple SiNx modes, as shown in
Fig.~\ref{fig:fig2}(B), where the blue and red curves represent the corresponding amplitude and phase response, respectively. Distinct phase transitions are observed at the frequencies corresponding to the coupling with specific SiNx modes. The blue curve in Fig.~\ref{fig:fig2}(C) displays a high resolution spectrum that
highlights the dense population of SiNx modes within the same frequency
range as the broad graphene resonance. The red curve in Fig.~\ref{fig:fig2}(C) shows the corresponding phase plot, exhibiting clear phase shifts associated with nearly every silicon nitride mode, demonstrating that the graphene resonator
interacts with a large number of these closely spaced modes.

\section*{Results and discussion}

\begin{figure*}[tb]
\includegraphics[width=.90\columnwidth]{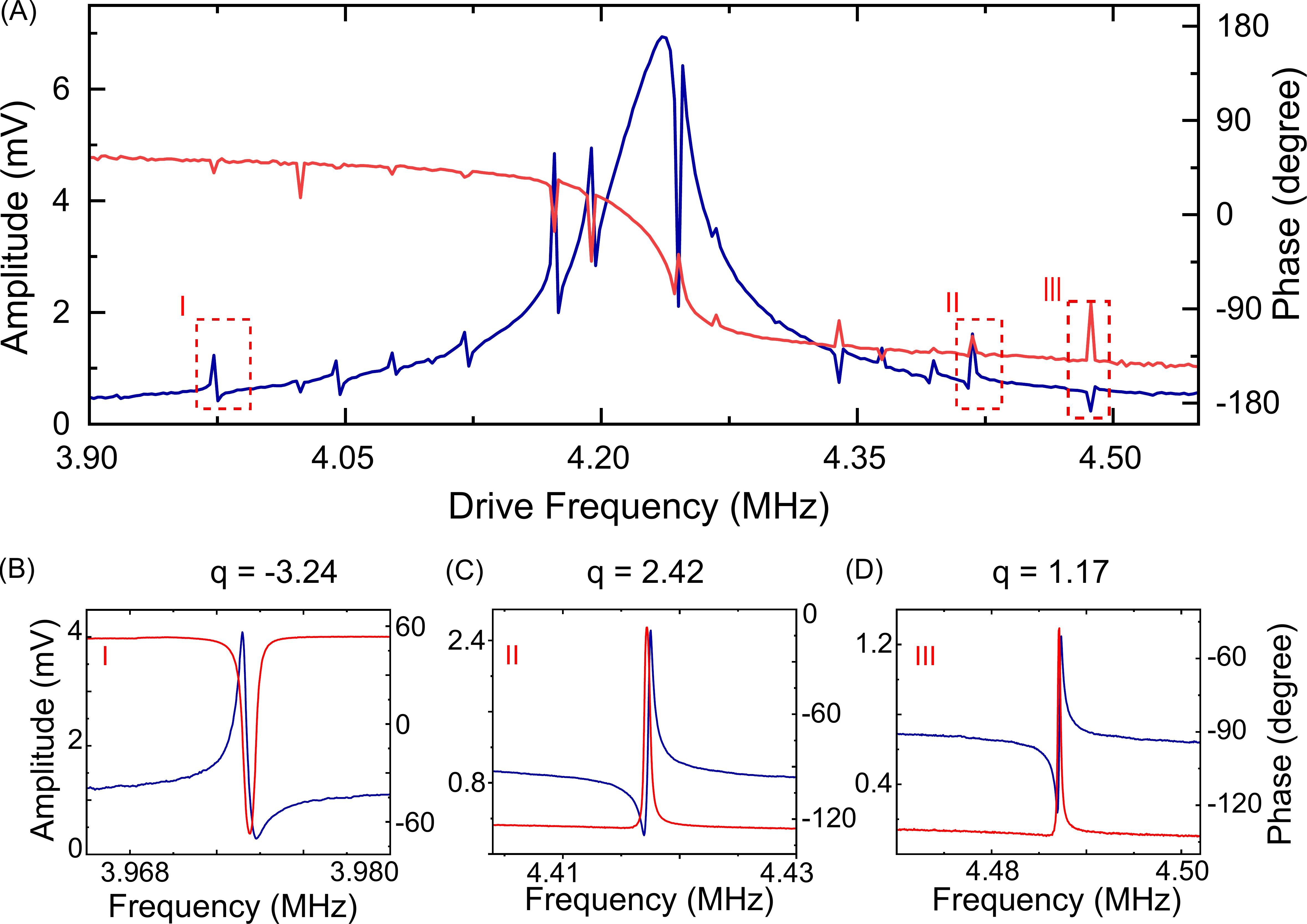}
\caption{\textbf{Fano resonances:} \textbf{(A)} Fano resonances with varying asymmetries are observed at different positions of the graphene mode. Four representative line shapes, labeled \textbf{I--III} and highlighted in red rectangles, are shown along with the corresponding phase responses of the graphene--SiNx hybrid modes (right axis). Panels \textbf{B--D} present the amplitude (upper panels) and phase (lower panels) responses for regions \textbf{I--III}. Regions \textbf{I}, exhibit negative Fano parameters, while regions \textbf{II} and \textbf{III} display positive Fano parameters, demonstrating the evolution of resonance asymmetry.}
\label{fig:fig3}
\end{figure*}

The resonance parameters of the graphene resonator were first determined by
fitting a Lorentzian line shape to the measured spectra, resulting in a
resonance frequency $\omega_g/2\pi = 4.23~\mathrm{MHz}$ and quality
factor $Q_g \approx 47$ at $V_{dc}=86$~V. Under these conditions,
both the amplitude response of the resonators and the phase of the driven
motion relative to the drive were recorded. At this fixed gate voltage, the
frequency spectrum of the graphene resonator indicates coupling to multiple
silicon nitride modes, as shown in Fig.~\ref{fig:fig3}(A): the broad graphene
fundamental mode interacts simultaneously with several sharp,
high quality factor SiNx modes. To examine these interactions in detail,
three representative SiNx modes located within or near the graphene
resonance are selected (marked by red rectangular boxes and labeled I, II, and
III in Fig.~\ref{fig:fig3}(A)). The regions enclosed by the red boxes are shown at higher magnification in the subsequent panels to examine the asymmetric resonance line shape.

Mode~I, located on the red-detuned side of the graphene resonance peak,
exhibits a Fano line shape with negative asymmetry parameter ($q<0$): the
amplitude rises to a peak and then drops to a sharp dip as the frequency
increases, as shown in Fig.~\ref{fig:fig3}(B). Across the resonance, the phase exhibits the characteristic $\pi$ shift expected for a driven resonator. Modes~II and~III (Fig.~\ref{fig:fig3}(C,~D) respectively), situated on
the blue-detuned side of the graphene peak, display Fano line shapes with the
opposite asymmetry ($q>0$), resulting in distinctly different phase
evolution. The evolution of the Fano asymmetry shown in Fig.~\ref{fig:fig3} is discussed below, and in the Supporting Information (SM)~\cite{supplementFano}. This asymmetry is governed by the relative contributions of the discrete and continuum components, the quality factor, and the coupling to the individual SiNx mode.

In order to quantitatively analyze this behavior, we introduce a toy model that consists of two coupled harmonic oscillators, which represent the graphene resonator and a vibrational mode of the silicon nitride membrane, respectively. The oscillators possess masses $m_g$ and $m_s$, spring constants $k_g$ and $k_s$, and damping coefficients $c_g$ and
$c_s$. The uncoupled fundamental resonance frequencies are
$\omega_g/2\pi$ (with $k_g=m_g\omega_g^{2}$) and $\omega_s/2\pi$
(with $k_s=m_s\omega_s^{2}$), where $\omega_g$ and $\omega_s$ are the angular
frequencies of the graphene and silicon nitride oscillators, respectively.
The oscillators are mechanically coupled through a spring of coupling
constant $\alpha$. Unlike conventional mechanical systems, the detuning
between the resonators and thereby the interference is tunable via the
applied gate voltage $V_{dc}$, enabling dynamic control over the resonance
behavior of the coupled system. The graphene resonator is driven by the
harmonic force $F=F_0\cos(\omega t)=\mathrm{Re}[F_0e^{i\omega t}]$ of
Eq.~(\ref{eq:drive}). The equations of motion are
\begin{align}
m_g\ddot{x}_g+c_g\dot{x}_g+(k_g+\Delta k)x_g+\alpha(x_g-x_s)&=F,
\label{eq:eomg}\\
m_s\ddot{x}_s+c_s\dot{x}_s+k_sx_s+\alpha(x_s-x_g)&=0,
\label{eq:eoms}
\end{align}

\begin{figure*}[tb]
\includegraphics[width=1.0\textwidth]{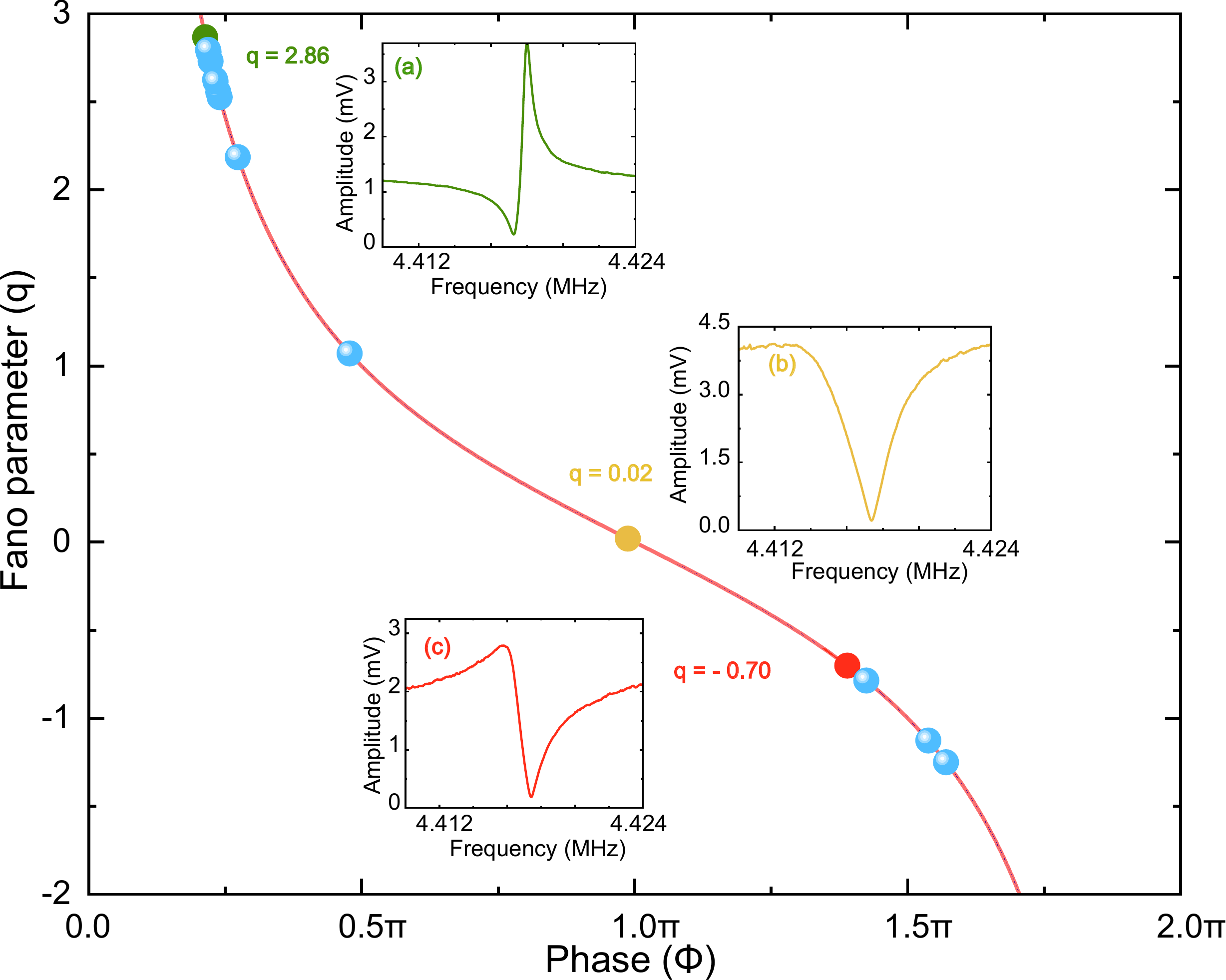}
\caption{\textbf{Evolution of the Fano asymmetry parameter.} The Fano asymmetry parameter, $q$, is shown as a function of gate voltage, $V_{dc}$, across an accumulated interference phase in a range of $2\pi$. The parameter $q$ transitions continuously from positive to negative values as the gate voltage increases. Representative data points at selected gate voltages are indicated. Insets (a)--(c) display the corresponding SiNx Fano line shapes: inset (a) (green) illustrates a positive Fano parameter ($q=2.86$), inset (b) (yellow) presents a nearly symmetric resonance with $q=0.02$, and inset (c) (blue) depicts a negative Fano parameter ($q=-0.70$).}
\label{fig:fig4}
\end{figure*}
where $x_g$ and $x_s$ denote the vibrational displacements of the graphene
and SiNx resonators. Graphene is a much lighter resonator than SiNx
(the mass ratio is of order $10^{4}$), and the two structures differ in
geometry and built-in tension, so their amplitude frequency responses are
very distinct. The stiffness of graphene is subject to an electrostatic
perturbation, $k_g \rightarrow k_g+\Delta k(V_{dc})$.
Fig.~\ref{fig:fig2}(A) demonstrates the shift of the graphene resonance
with increasing gate voltage and the evolution of its interaction with a
fixed (shaded) silicon nitride mode: as the graphene mode is tuned across the
SiNx resonance, characteristic Fano interference features emerge in the
spectrum.

Solving Eqs.~(\ref{eq:eomg})--(\ref{eq:eoms}) in the steady state and
expanding near the sharp SiNx resonance yields the standard Fano
intensity profile (SM~\cite{supplementFano}, Sec.~II),
\begin{equation}
\left|\frac{x_g(\omega)}{x_{\rm bg}(\omega)}\right|^{2}
=\frac{(\epsilon+q)^{2}+b^{2}}{\epsilon^{2}+1},
\qquad
q=-\cot\phi ,
\label{eq:fano}
\end{equation}
where $x_{\rm bg}$ is the smooth background response of the bare graphene
mode, $\epsilon$ is the drive detuning from the (weakly shifted) SiNx
resonance normalized by the half linewidth $\Gamma/2$, and
$\phi=\arg[\,\omega_g'^{2}-\omega_s^{2}+i\gamma_g\omega_s\,]$ is the phase of
the continuum (graphene) response evaluated at the discrete mode frequency,
with $\omega_g'$ the gate shifted graphene frequency and
$\gamma_g=c_g/m_g$. The Fano parameter depends on the geometric arrangement, material properties, excitation conditions, and, for this platform, the gate voltage. As $V_{dc}$ is varied, the relative detuning changes and $\phi$ evolves continuously, resulting in a continuous variation of $q$. The theoretical framework also predicts a coupling induced, Purcell like broadening of the SiNx mode, $\Gamma=\gamma_s(1+C)$, and a dip floor $b=1/(1+C)$ determined by the cooperativity. $C=\lambda^{2}\gamma_g/(\gamma_s|d_g|^{2})$, where $\lambda^{2}=\alpha^{2}/(m_gm_s)$ and $d_g$ is the complex graphene response denominator (SM~\cite{supplementFano}). The value and sign of $q$ directly determine the
characteristic Fano line shape, which ranges from asymmetric dips and peaks
to nearly symmetric Lorentzian like profiles, providing insight into the
strength and nature of the mode coupling in the hybrid graphene SiNx
system. Numerical solutions of the full coupled mode model with parameters
representative of the experiment quantitatively reproduce the line shapes and
phase jumps of Fig.~\ref{fig:fig3} (SM~\cite{supplementFano}, Sec.~IV).

Fig.~\ref{fig:fig4} presents the measured evolution of the Fano parameter $q$, which spans a total accumulated interference phase in a range of $2\pi$ (SM~\cite{supplementFano}, Sec.~V). At the lowest gate voltages, the Fano parameter remains nearly constant due to the relatively weak electrostatic tuning of the graphene resonance. This results in only minor changes in the detuning between the graphene and SiNx modes. As the gate voltage $V_{dc}$ increases, the Fano parameter continuously decreases from its maximum value, crosses zero, and eventually reaches a minimum value of $q=-1.25$.
At a gate voltage of $V_{dc}=106$~V, the Fano parameter reaches its maximum value of $q=2.86$, represented by the green data point, and the corresponding Fano line shape is shown in Fig.~\ref{fig:fig4}, inset (a). With further increases in gate voltage, the graphene resonance approaches that of the SiNx mode, resulting in a nearly symmetric absorption peak as the Fano parameter approaches zero, indicated by the yellow data point in Fig.~\ref{fig:fig4}. The corresponding line shape appears in Fig.~\ref{fig:fig4}, inset (b). Upon additional increases in gate voltage, the resonance asymmetry reverses due to a phase change. As a result, the Fano parameter becomes negative, represented by the blue data point, with the corresponding line shape shown in Fig.~\ref{fig:fig4}, inset (c).
These results demonstrate that the gate voltage enables direct control over both the sign and magnitude of the Fano asymmetry parameter, allowing continuous tuning of the characteristic Fano line shape from asymmetric peaks to nearly symmetric resonances and subsequently to asymmetric dips. This voltage controlled Fano resonance positions the hybrid graphene SiNx resonator as a promising platform for highly tunable nanomechanical devices, particularly in high sensitivity sensing applications.
\\Previous studies have demonstrated Fano resonances in systems such as coupled
silicon micromechanical resonators~\cite{Xie2025} and micro cantilever
arrays~\cite{Stassi2017}. In coupled silicon micromechanical resonators, the
resonance frequency and coupling strength are tunable to a limited degree;
however, these systems are generally restricted to interactions between two
resonators and offer limited control over the Fano asymmetry parameter, so
the range of accessible Fano line shapes including transitions between
positive and negative asymmetry is constrained. Micro cantilever arrays, in
turn, have fixed geometrical and mechanical parameters set during
fabrication, allowing minimal or no post fabrication tunability. The present
hybrid graphene SiNx platform instead consists of multiple coupled
SiNx modes interacting with a broadly tunable graphene resonator. Owing to
graphene's electrostatic tunability, its resonance frequency can be
continuously adjusted, enabling dynamic control over the interactions between
resonant modes. This approach enables precise tuning of the Fano line shape
and the corresponding asymmetry parameter across a wide range, and provides a
unique opportunity to systematically investigate the relationship between the
Fano parameter and the phase difference that gives rise to the asymmetric
resonance profile.

\section*{Frequency locking in the nonlinear regime}

\begin{figure*}[!ht]
    \centering
    \includegraphics[width=0.95\textwidth]{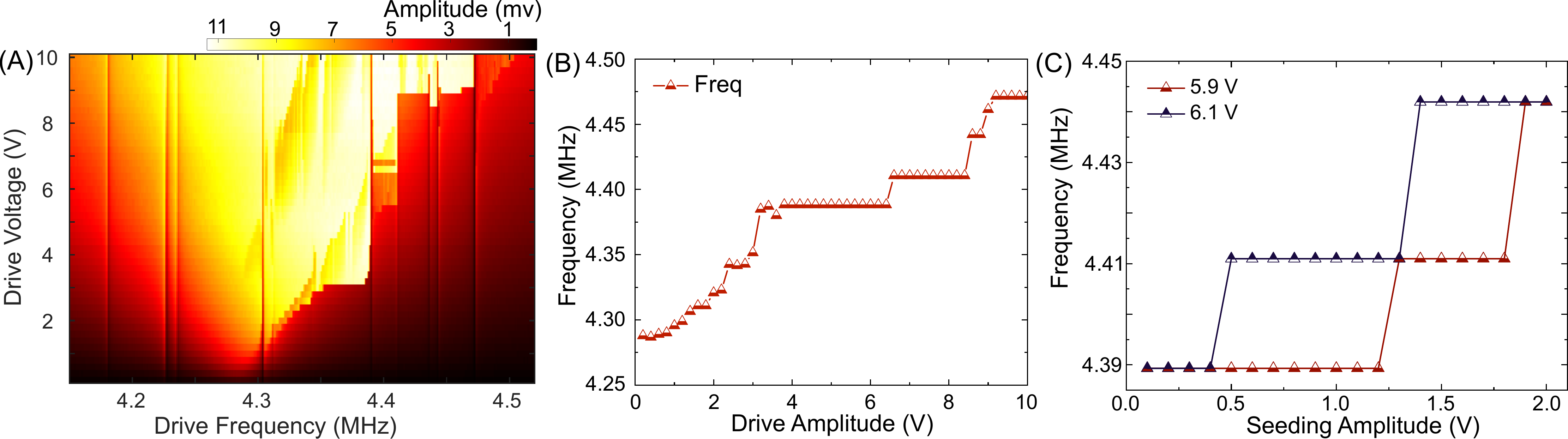}
        \caption{\textbf{Frequency locking of the nonlinear graphene mode to the SiNx mode comb at $V_{dc} = 113$~V.} \textbf{(A)} Measured graphene amplitude as a function of drive frequency and drive voltage amplitude $V_{ac}$. The antiresonance lines remain pinned at the fixed SiNx frequencies at all drive levels, and the jump-down edge of the hardening Duffing response forms a staircase locked to successive SiNx modes. \textbf{(B)} Jump-down frequency $\omega_J/2\pi$ versus drive amplitude, extracted from (A): $\omega_J/2\pi$ is pinned on plateaus at the SiNx frequencies, within which $\partial (\omega_J/2\pi)/\partial V_{ac} \approx 0$ (frequency stabilization); the plateau widths reflect the coupling strength (cooperativity) of the individual modes. \textbf{(C)} Deterministic switching between locked states: with the primary drive held on a plateau ($V_{ac} = 5.9$~V and $6.1$~V), a weak seeding tone applied at the locking point steps $\omega_J/2\pi$ between adjacent plateaus ($4.389 \rightarrow 4.410 \rightarrow 4.442$~MHz) at sharp threshold seeding amplitudes that decrease with increasing primary drive.}
\label{fig:fig5}
\end{figure*}
The results discussed so far were obtained at weak AC drive, where the
graphene resonator responds linearly. Owing to its atomic thickness and
membrane geometry, graphene acquires a hardening (Duffing) nonlinearity at
very moderate drive amplitudes~\cite{Chen2009,Eichler2011}: the equation of
motion of the graphene coordinate in Eq.~(\ref{eq:eomg}) acquires a cubic
restoring term $\beta x_g^{3}$, the response becomes bistable, and an
upward frequency sweep follows the high amplitude branch until it collapses
at the saddle node bifurcation. In an isolated Duffing resonator the
jump-down frequency $\omega_J/2\pi$ advances {continuously} with drive amplitude,
and 
%amplitude fluctuations are converted into frequency fluctuations
amplitude noise is transduced into
noise of the oscillation frequency~\cite{Antonio2012}.

In the hybrid device the behavior is strikingly different.
Fig.~\ref{fig:fig5}(A) maps the measured graphene amplitude as a
function of drive frequency and drive voltage amplitude $V_{ac}$
(0.2 -- 10~V) at a fixed gate voltage $V_{dc}=113$~V. Two features
stand out. First, the dark vertical antiresonance lines pinned at the fixed
SiNx frequencies persist at all drive levels: this is expected from the
coupled oscillator theory, in which the response zero is locked to the
discrete mode frequency independently of the graphene parameters
(SM~\cite{supplementFano}, Sec.~I), and it remains true in the nonlinear regime. Second,
the jump-down edge of the Duffing response does not advance smoothly with
$V_{ac}$; instead it remains pinned at a SiNx frequency over an extended
range of drive and then snaps abruptly to the next mode, so that the
bistability boundary forms a staircase across the map.

Fig.~\ref{fig:fig5}(B) shows the jump-down frequency $\omega_J/2\pi$ extracted
from these sweeps as a function of $V_{ac}$. Rather than the smooth growth of
an isolated Duffing mode, $\omega_J/2\pi$ exhibits a staircase of plateaus at
4.342, 4.389, 4.410, 4.442, and 4.471~MHz, coinciding with the
antiresonance frequencies of the SiNx modes in
Fig.~\ref{fig:fig5}(A). On the widest plateau the jump frequency remains
fixed at 4.389~MHz while the drive voltage nearly doubles (from
3.8~V to 6.4~V); within a locked plateau the sensitivity
$\partial (\omega_J/2\pi)/\partial V_{ac}$ is suppressed to nearly zero, i.e., the
oscillation frequency is stabilized against drive and amplitude
fluctuations. The mechanism follows directly from the Fano physics
established above: as the drive pushes the high amplitude Duffing branch
toward a SiNx mode, the pinned antiresonance drains energy from the
graphene motion into the high-$Q$ discrete mode and destructively suppresses
the graphene response, so the upper branch cannot be sustained past the
interference zero and the jump-down is triggered at the SiNx frequency.
Only when the drive becomes strong enough for the branch to traverse the
antiresonance dip does $\omega_J/2\pi$ snap to the next mode of the comb. The locking
range of each plateau is accordingly set by the depth of the corresponding
dip, $b=1/(1+C)$ of Eq.~(\ref{eq:fano}), i.e., by the cooperativity of that
particular SiNx mode; the markedly unequal plateau widths in
Fig.~\ref{fig:fig5}(B) thus map the distribution of linear and nonlinear
coupling strengths across the mode comb. Numerical sweeps of the
Duffing extended coupled mode model reproduce both the pinned staircase and
the plateau width hierarchy (SM~\cite{supplementFano}, Sec.~VIII). This realizes, in a single
two terminal device, the frequency stabilization demonstrated for internal
resonances in micromechanical oscillators~\cite{Antonio2012}, with the
important difference that the dense SiNx comb provides an array of
stable, lithographically defined set points to which the graphene oscillator
can be locked.

The locked states can moreover be addressed deterministically.
Fig.~\ref{fig:fig5}(C) shows the response when, with the primary drive
held fixed on a plateau, a second (seeding) tone is applied at the
locking point and its amplitude is ramped. The locked frequency does not
shift continuously: it remains on the initial plateau
(4.389~MHz) up to a sharp threshold seeding amplitude, then steps
discretely to the next plateau (4.410~MHz) and subsequently to the
following one (4.442~MHz). The switching thresholds shift
systematically to lower seeding amplitude when the primary drive is
increased from 5.9~V to 6.1~V, consistent with the
seed supplying the residual energy needed to carry the Duffing branch across
the antiresonance barrier. The hybrid resonator therefore behaves as a
multi-stable frequency element whose state is selected digitally by a weak
control tone a mechanical analog of injection locked frequency
switching~\cite{Houri2017} suggesting applications in frequency-shift
keying, mechanical multi-level memories, and reconfigurable clocking of
NEMS based signal processors.

\section*{Conclusion}
A tunable Fano resonance was experimentally demonstrated in a hybrid
graphene silicon nitride (SiNx) nanomechanical system. The observed
asymmetric resonance results from interference between a broad graphene
resonance and a narrow SiNx resonance, forming a mechanical analog of the
Fano effect in coupled oscillators. Unlike conventional Lorentzian
resonances, Fano resonances exhibit asymmetric line shapes with
significantly steeper spectral responses, which are advantageous for sensing
and signal processing applications. Adjusting the gate voltage alters the
tension in the graphene membrane, thereby shifting its frequency and enabling
precise control over the interaction between graphene and SiNx;
consequently, the Fano asymmetry parameter $q$ can be continuously tuned
across a wide range. The measurements reveal a direct relationship between
the Fano parameter and the phase difference of the interfering resonant
pathways, $q=-\cot\phi$, offering experimental and analytical insight into
the origin of the Fano asymmetry. Driving the same device into the nonlinear
regime uncovered a second consequence of the hybrid coupling: the jump-down
frequency of the graphene Duffing mode locks to the fixed SiNx mode comb,
producing drive insensitive frequency plateaus that stabilize the oscillation
against amplitude noise, and a weak seeding tone switches the resonator
deterministically between adjacent locked states. The same
interference the antiresonance pinned at each SiNx frequency thus
governs both regimes, shaping the linear response into tunable Fano profiles
and quantizing the nonlinear response into a staircase of stable frequency
states. This demonstrated control over the resonance line shape, the Fano
parameter, and the locked oscillation frequency underscores the potential of
hybrid graphene based nanomechanical systems for tunable sensors, stable
frequency references, mechanical switches and memories, and reconfigurable
signal processing devices.

%\section*{acknowledgments}
\begin{acknowledgments}
This work is supported by the National Quantum Mission of DST, ERC Grant
No.~639739, and DFG Project GZ: BO 5142/4-1, the Priority Program SPP 2244, the German Excellence Strategy - EXC3112/1 - 533767171 (Center for Chiral Electronics), and the Federal Ministry of Education and Research (BMBF, project 05K22KE3). A.K.R. acknowledges MHRD for financial support.
J.A.M. acknowledges the Prime Minister's Research Fellows (PMRF) scheme of
the Ministry of Human Resource Development, Govt.\ of India, for financial
support.
\end{acknowledgments}

\bibliography{Fano}

\newpage

\title{Tunable Fano Resonance and Frequency Locking in a Graphene-SiNx Hybrid
Nanomechanical Resonator}
\maketitle

\setcounter{equation}{0}
\renewcommand{\theequation}{S.\arabic{equation}}
\renewcommand\thefigure{S.\arabic{figure}}    
\setcounter{figure}{0} 
\makeatletter

\section{Coupled-oscillator model and exact steady state}
\label{sec:model}

We model the hybrid device as one driven, heavily damped oscillator (the
graphene fundamental mode, playing the role of the {continuum}) coupled
by weak springs to $N$ high-quality-factor oscillators (the SiN$_x$ modes,
playing the role of the {discrete states}). An intuitive picture is a
muffled drum tied by thin threads to a set of tuning forks: only the drum is
struck, but whenever the driving tone matches a fork, the fork rings up and
pushes back on the drum, and the drum's motion is the interference of the two
pathways (direct drive, and drive routed through the fork). It is this
two-pathway interference that produces the Fano line shape.

The equations of motion [Eqs.(~\ref{eq:eomg})--(\ref{eq:eoms}) of the main text, generalized to $N$ discrete modes] read
\begin{align}
m_g\ddot{x}_g+c_g\dot{x}_g+(k_g+\Delta k)\,x_g
+\sum_{n=1}^{N}\alpha_n\,(x_g-x_n)&=F_0\cos(\omega t),
\label{eq:eomg_g}\\
m_n\ddot{x}_n+c_n\dot{x}_n+k_nx_n+\alpha_n\,(x_n-x_g)&=0,
\qquad n=1,\dots,N .
\label{eq:eomn}
\end{align}
Writing $x_j(t)=\mathrm{Re}[X_j\,e^{i\omega t}]$ and dividing by the masses,
the steady-state amplitudes obey
\begin{align}
\left[d_g(\omega)\right]X_g-\sum_n \frac{\alpha_n}{m_g}X_n &=\frac{F_0}{m_g},
&
\left[d_n(\omega)\right]X_n&=\frac{\alpha_n}{m_n}X_g ,
\label{eq:linear}
\end{align}
with the {dressed} response denominators
\begin{equation}
d_g(\omega)=\omega_g'^{2}-\omega^{2}+i\gamma_g\omega,
\qquad
d_n(\omega)=\omega_n'^{2}-\omega^{2}+i\gamma_n\omega,
\label{eq:denoms}
\end{equation}
where $\omega_g'^{2}=(k_g+\Delta k+\sum_n\alpha_n)/m_g$,
$\omega_n'^{2}=(k_n+\alpha_n)/m_n$, $\gamma_g=c_g/m_g$, and
$\gamma_n=c_n/m_n$. Eliminating the $X_n$ gives the exact driven response of
the graphene mode,
\begin{equation}
\boxed{\;
X_g(\omega)=\frac{F_0/m_g}{\,d_g(\omega)-\Sigma(\omega)\,},
\qquad
\Sigma(\omega)=\sum_{n}\frac{\lambda_n^{2}}{d_n(\omega)},
\qquad
\lambda_n^{2}\equiv\frac{\alpha_n^{2}}{m_g m_n}\; ,}
\label{eq:exact}
\end{equation}
and $X_n=(\alpha_n/m_n)\,X_g/d_n(\omega)$. The quantity $\Sigma(\omega)$ is a
mechanical {self-energy}: each sharp mode ``pushes back'' on the
graphene with a strength that diverges on its own resonance. Because the mass
ratio is $m_n/m_g\sim10^{4}$, physically reasonable coupling springs give
$\lambda_n^{2}$ small compared with $|d_g|$ except within the narrow SiN$_x$
linewidthsn this is the weak-coupling (Fano, rather than normal-mode
splitting) regime of the experiment.

Two exact statements follow directly from Eq.~(\ref{eq:exact}) for a single
discrete mode ($N=1$) with negligible intrinsic loss ($\gamma_s\to0$):
(i)~$X_g$ has an exact zero (perfect antiresonance) at
$\omega^{2}=\omega_s'^{2}=(k_s+\alpha)/m_s$, i.e., the dip is pinned to the
discrete-mode frequency, independent of all graphene parameters; and
(ii)~$X_g$ has a pole (peak) displaced from the zero by an amount controlled
by the coupling. The asymmetric peak dip pair separated by a small,
coupling controlled interval is the anatomy of the Fano line.

\section{Analytic Fano reduction: {$q=-\cot\phi$}, induced linewidth, and contrast}
\label{sec:fano}
%\label{main:Fano}

Consider a single discrete mode and normalize the response to the smooth
background $X_{\rm bg}(\omega)=(F_0/m_g)/d_g(\omega)$ of the bare graphene
mode:
\begin{equation}
R(\omega)\equiv\frac{X_g}{X_{\rm bg}}
=\frac{d_s(\omega)}{d_s(\omega)-\Sigma(\omega)},
\qquad \Sigma(\omega)=\frac{\lambda^{2}}{d_g(\omega)} .
\label{eq:ratio}
\end{equation}
Within the narrow SiN$_x$ linewidth, $d_g$ (and hence $\Sigma$) is constant
to excellent accuracy. Write it in polar form at the discrete-mode frequency,
\begin{equation}
d_g(\omega_s')=|d_g|\,e^{i\phi},
\qquad
\phi=\arg\!\left[\omega_g'^{2}-\omega_s'^{2}+i\gamma_g\omega_s'\right]
\in(0,\pi),
\label{eq:phi}
\end{equation}
so that $\Sigma=(\lambda^{2}/|d_g|)(\cos\phi-i\sin\phi)$. The phase $\phi$ is
the lag of the continuum: $\phi\to0$ when the discrete mode sits far on the
red side of the graphene resonance ($\omega_s'\ll\omega_g'$), $\phi=\pi/2$
exactly on resonance, and $\phi\to\pi$ far on the blue side.
\begin{figure}[!ht]
\includegraphics[width=0.86\textwidth]{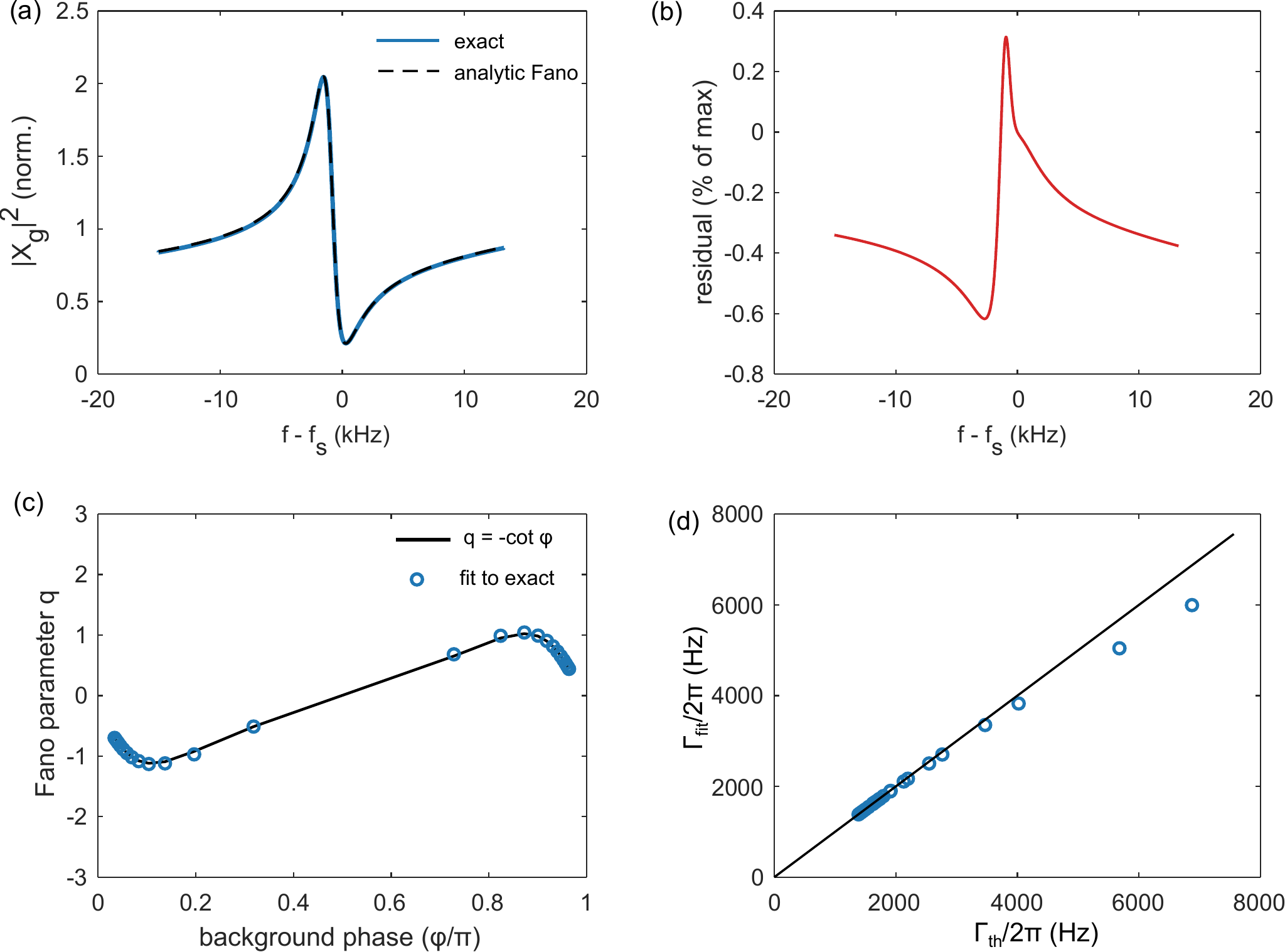}
\caption{Numerical validation of the analytic Fano reduction.
(a)~Exact coupled-oscillator response $|X_g|^{2}$ (blue) and the analytic
profile of Eqs.~(\ref{eq:qresult})--(\ref{eq:fanoprofile}) (dashed black) for
a discrete mode at 4.05~MHz, red-detuned from the graphene resonance at
4.23~MHz. (b)~Residual between the two curves. (c)~Fano parameter extracted
by fitting the exact response (circles) versus the analytic prediction
$q=-\cot\phi$ (line) as the discrete-mode frequency is swept across the
graphene resonance. (d)~Fitted total linewidth versus the analytic
$\Gamma=\gamma_s(1+C)$ of Eq.~(\ref{eq:purcell}).}
\label{fig:S1}
\end{figure}
Linearizing $d_s$ around the discrete resonance, with
$\delta=\omega-\omega_s'$,
$d_s\simeq-2\omega_s'\left(\delta-i\gamma_s/2\right)$, Eq.~(\ref{eq:ratio})
becomes
\begin{equation}
R=\frac{\delta-i\gamma_s/2}{(\delta-\delta_0)-i\Gamma/2},
\qquad
\delta_0=-\frac{\lambda^{2}\cos\phi}{2\omega_s'|d_g|},
\qquad
\frac{\Gamma}{2}=\frac{\gamma_s}{2}+\frac{\lambda^{2}\sin\phi}{2\omega_s'|d_g|} .
\label{eq:linearized}
\end{equation}
The coupling therefore (i)~\emph{pulls} the resonance by $\delta_0$ and
(ii)~\emph{broadens} it by the induced rate
\begin{equation}
\Gamma_{\rm ind}
=\frac{\lambda^{2}\sin\phi}{\omega_s'|d_g|}
=\frac{\lambda^{2}\gamma_g}{|d_g(\omega_s')|^{2}}\;,
\qquad
\Gamma=\gamma_s+\Gamma_{\rm ind}=\gamma_s\,(1+C),
\qquad
C\equiv\frac{\Gamma_{\rm ind}}{\gamma_s},
\label{eq:purcell}
\end{equation}
where we used $\sin\phi=\gamma_g\omega_s'/|d_g|$. This is the exact
mechanical analog of Purcell broadening in cavity QED: the sharp SiN$_x$
mode (the ``atom'') leaks energy through the lossy graphene mode (the
``cavity''), and $C$ is the familiar cooperativity. Defining the normalized
detuning $\epsilon=(\delta-\delta_0)/(\Gamma/2)$ and substituting into
Eq.~(\ref{eq:linearized}),
\begin{equation}
R(\epsilon)=\frac{\epsilon+q-ib}{\epsilon-i},
\qquad
\boxed{\;q=\frac{2\delta_0}{\Gamma}\;\xrightarrow[\;\gamma_s\to0\;]{}\;
-\frac{\cos\phi}{\sin\phi}=-\cot\phi\;},
\qquad
b=\frac{\gamma_s}{\Gamma}=\frac{1}{1+C},
\label{eq:qresult}
\end{equation}
so that the measured intensity takes the generalized Fano form
\begin{equation}
\left|R(\epsilon)\right|^{2}
=\frac{(\epsilon+q)^{2}+b^{2}}{\epsilon^{2}+1}\; .
\label{eq:fanoprofile}
\end{equation}
Equations~(\ref{eq:qresult})--(\ref{eq:fanoprofile}) are the central analytic
results. They make three parameter-free predictions:

\begin{enumerate}
\item \textbf{Sign of the asymmetry.} On the red-detuned side of the graphene
resonance ($\omega_s'<\omega_g'$), $\mathrm{Re}\,d_g>0$ so $\phi<\pi/2$ and
$q<0$: the line rises to a peak and falls to a dip with increasing frequency.
On the blue-detuned side, $\phi>\pi/2$ and $q>0$: dip first, then peak. This
is precisely the pattern of modes I (red, $q<0$) versus II and III (blue,
$q>0$) in Fig.~3 of the main text.
\item \textbf{Exactly on resonance} ($\phi=\pi/2$), $q=0$: a pure,
symmetric antiresonance dip.
\item \textbf{Dip contrast.} The dip floor is $b^{2}/(1)$ at $\epsilon=-q$:
the antiresonance reaches zero only in the limit of large cooperativity
$C\gg1$, i.e., when the coupling-induced decay dominates the intrinsic
SiN$_x$ loss. Finite dip depths in the data directly measure $C$.
\end{enumerate}

The line-shape extrema follow from $\partial_\epsilon|R|^2=0$ (for $b=0$):
the minimum ($|R|^2=0$) lies at $\epsilon=-q$ and the maximum
($|R|^{2}=1+q^{2}$) at $\epsilon=1/q$, so the peak-to-dip frequency
separation is
\begin{equation}
\Delta\omega_{\rm pk-dip}=\frac{\Gamma}{2}\left|q+\frac{1}{q}\right|\ge\Gamma,
\label{eq:pkdip}
\end{equation}
minimized at $|q|=1$, where the profile is maximally ``dispersive.''

\section{Numerical validation of the analytic results}
\label{sec:validation}

Because the reduction of  Sec.~\ref{sec:fano} 
involves a linearization, we
verified it against the {exact} response Eq.~(\ref{eq:exact}), computed
numerically with parameters representative of the experiment
($\omega_g/2\pi=4.23$~MHz, $Q_g=50$, $Q_s=3\times10^{3}$; couplings $\lambda^{2}$
chosen to give induced linewidths of a few kHz, as observed). The comparison
is shown in Fig.~\ref{fig:S1}. Panel (a) overlays the exact and analytic
profiles for a representative red-detuned mode; panel (b) shows the residual,
which remains below ${\sim}1\%$ of the peak and is concentrated at the sharp
extremum, where it reflects a small $\mathcal{O}(\lambda^{4})$ correction to
the pole position rather than any deviation of the line {shape}: the
extracted shape parameters agree far better, as panels (c) and (d) show. Panels (c) and
(d) sweep the discrete-mode frequency across the entire graphene resonance:
the Fano parameter obtained by {fitting} Eq.~(\ref{eq:fanoprofile}) to
the exact curve traces the analytic law $q=-\cot\phi$, and the fitted
linewidth reproduces $\Gamma$ of Eq.~(\ref{eq:purcell}), both at the percent
level. The analytic expressions can therefore be used to invert measured
line shapes for the physical parameters $(\phi,\lambda^{2},C)$.

\section{Reproducing the experimental signatures of Fig.~3: modes I--III}
\label{sec:modes}
\begin{figure}[!htb]
\includegraphics[width=1.0\textwidth]{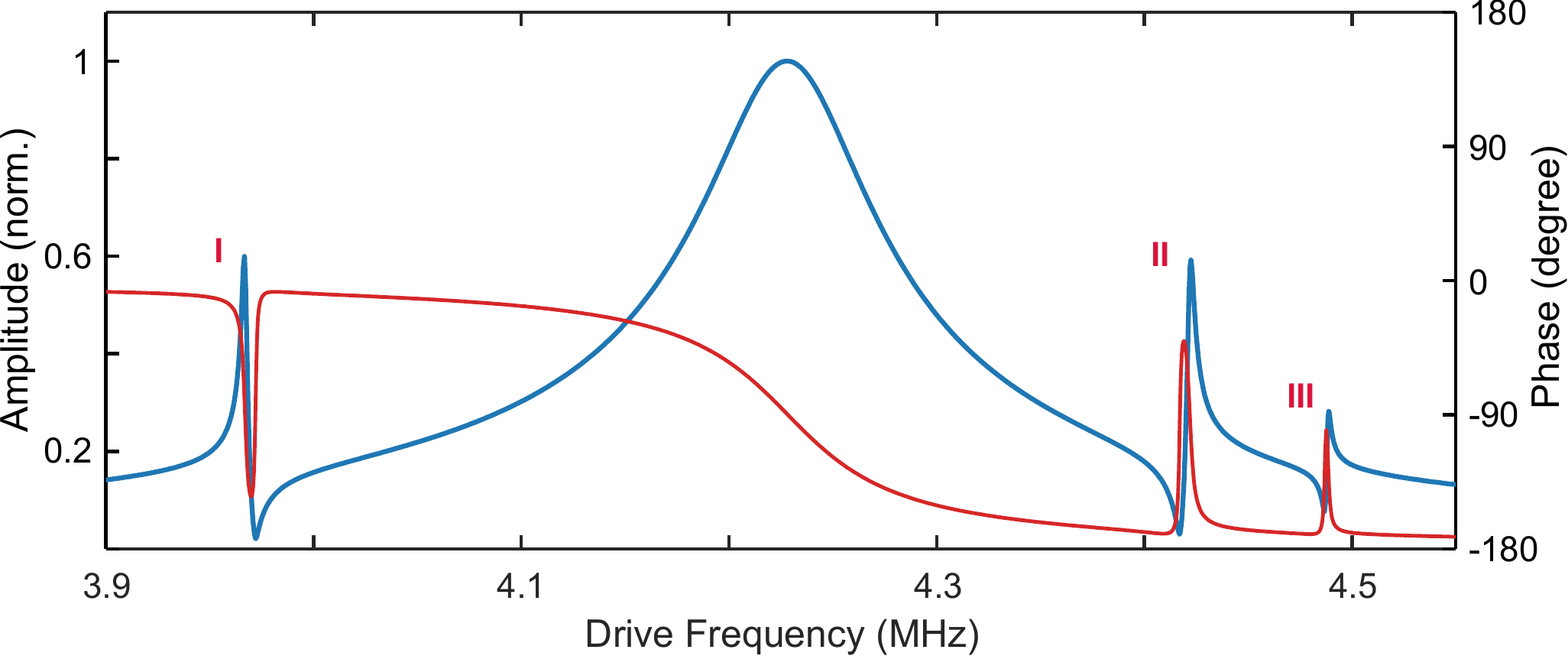}
\caption{Simulated wide-band amplitude spectrum of the detected signal for
the three-mode model at $\omega_g/2\pi=4.23$~MHz, to be compared with the top panel of
Fig.~3 of the main text: a broad graphene resonance decorated with sharp,
asymmetric Fano features at the SiN$_x$ mode frequencies (labels I--III).}
\label{fig:S2wide}
\end{figure}
We next simulate the full multi-mode system with the three modes highlighted
in Fig.~3 of the main text ($\omega_s/2\pi=3.972$, $4.417$, and $4.487$~MHz)
coupled simultaneously to the graphene mode at $\omega_g/2\pi=4.23$~MHz. In the
multi-mode case, the ``continuum'' seen by mode $n$ is the graphene response
renormalized by the tails of all other discrete modes; the asymmetry law
generalizes to
\begin{equation}
q_n=-\cot\phi_n,
\qquad
\phi_n=\arg\Big[\,d_g(\omega_n')-\sum_{m\neq n}
\frac{\lambda_m^{2}}{d_m(\omega_n')}\,\Big].
\label{eq:multiq}
\end{equation}
\begin{figure}[!htb]
\includegraphics[width=1.0\textwidth]{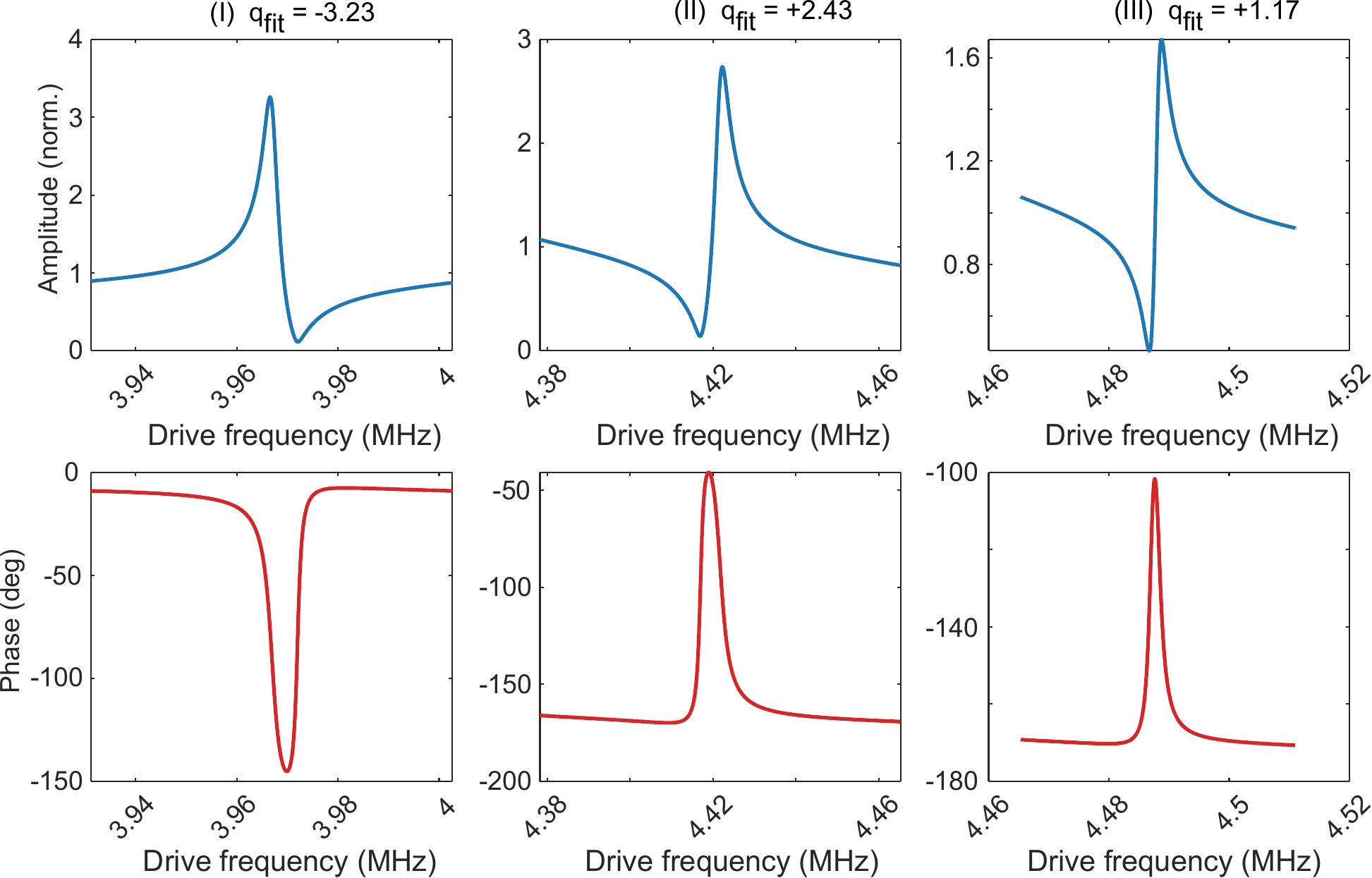}
\caption{Simulated amplitude (top row) and phase (bottom row) of the detected
signal near modes I--III. The
fitted Fano parameters reproduce the experimental sign pattern of Fig.~3 of
the main text: mode I peak-then-dip ($q<0$), modes II and III
dip-then-peak ($q>0$).}
\label{fig:S2}
\end{figure}
For the weak couplings relevant here, the correction from neighboring modes
is small, and modes I--III follow the two-oscillator prediction: mode~I
(red-detuned, $\phi<\pi/2$) has $q<0$; modes II and III (blue-detuned,
$\phi>\pi/2$) have $q>0$.

Figure~\ref{fig:S2wide} shows the simulated wide spectrum a broad graphene
peak decorated by sharp Fano features, closely resembling the top panel of
Fig.~3 of the main text and Fig.~\ref{fig:S2} zooms into the three modes.
The fitted asymmetries are $q_{\rm I}=-3.23$, $q_{\rm II}=+2.43$, and
$q_{\rm III}=+1.17$, matching the analytic predictions
of Eq.~(\ref{eq:multiq}) to within a few percent and
reproducing the experimental sign pattern $(-,+,+)$. The lower row shows the
corresponding simulated phase: each Fano feature is accompanied by a sharp
phase excursion of order $\pi$ across the induced linewidth, whose direction
correlates with the sign of $q$ the behavior seen in Figs.~$2$(B), $2$(C), and
$3$ of the main text.

\section{The $q$--$\Phi$ master curve of Fig.~4}
\label{sec:master}

As the gate voltage tunes the graphene frequency across a given SiN$_x$
mode, the continuum phase $\phi$ of Eq.~(\ref{eq:phi}) evolves from $\pi$
(graphene far below the mode) through $\pi/2$ (on resonance) to $0$
(graphene far above), and $q=-\cot\phi$ correspondingly sweeps from large
positive values through zero to large negative values, as seen
in Fig.~4 of the main text. It is convenient to define the
{accumulated} Fano phase $\Phi=2(\pi-\phi)$, which grows monotonically
with gate voltage and advance by $2\pi$ for each complete mode crossing
(each pathway discrete and continuum contributes a $\pi$ winding). In
this variable the asymmetry law reads
\begin{equation}
q=\cot(\Phi/2),
\label{eq:master}
\end{equation}
a universal, periodic ``master curve'': successive $2\pi$ window of $\Phi$
correspond to successive SiN$_x$ mode crossings and retrace the same
decreasing branch, from $q\to+\infty$ at the start of a window to
$q\to-\infty$ at its end.

\begin{figure}[tb]
\includegraphics[width=0.8\textwidth]{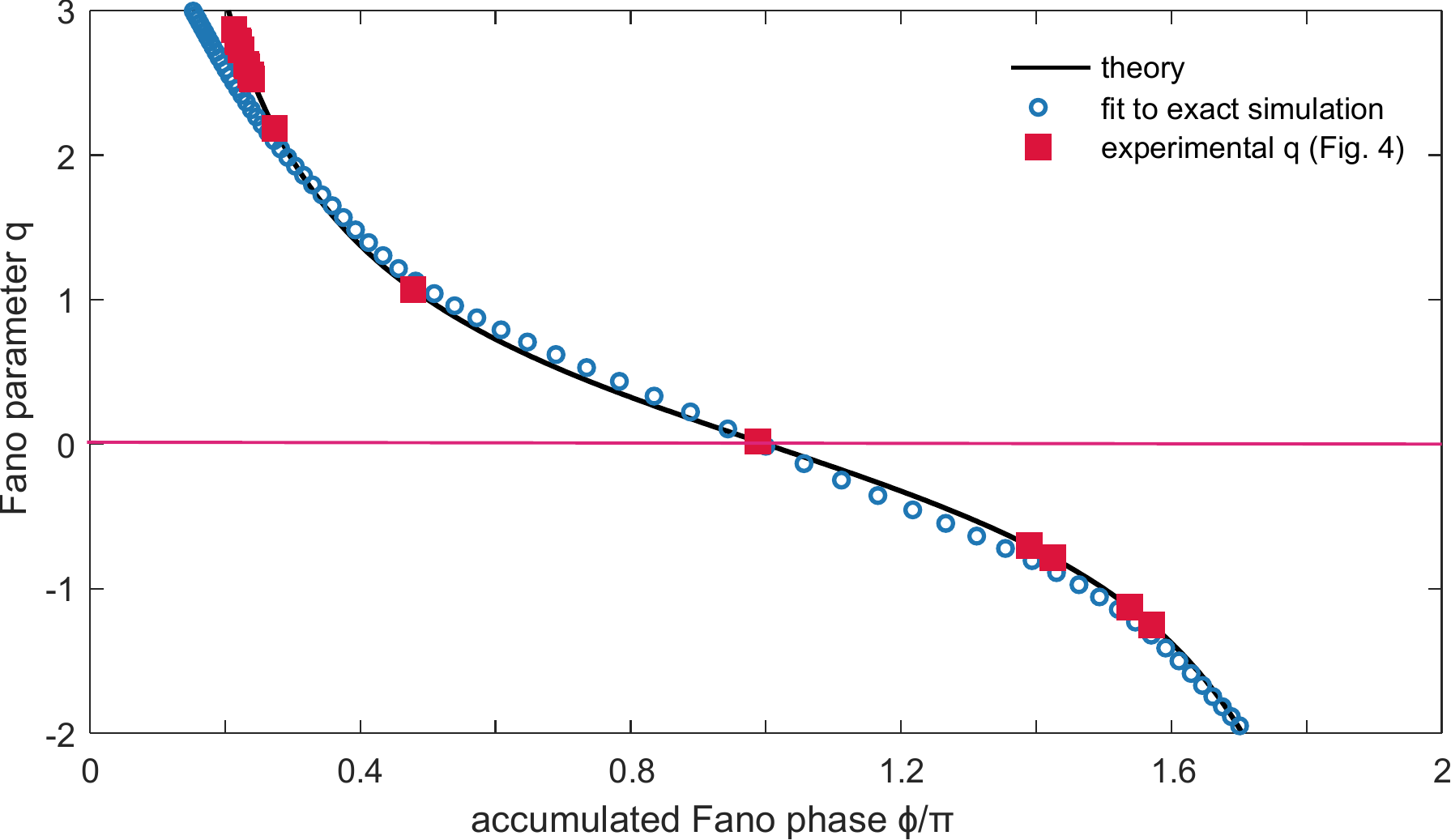}
\caption{The $q$--$\Phi$ master curve, Eq.~(\ref{eq:master}), over three
$2\pi$ window (Fig.~4
of the main text). Open circles: $q$ obtained by fitting exact simulated
spectra while the graphene frequency is tuned across a SiN$_x$ mode,
emulating the $V_{dc}$ sweep. Red squares: the experimental $q$ values of
Fig.~4, placed on the curve.}
\label{fig:S3}
\end{figure}

Fig.~\ref{fig:S3} show this master curve over a window
($0\le\Phi\le2\pi$), as shown in Fig.~4 of the main text. The open
circles are asymmetries obtained by fitting Eq.~(\ref{eq:fanoprofile}) to
exact simulated spectra as the graphene frequency is swept across a mode at
4.50~MHz (the numerical analog of the $V_{dc}$ sweep); they trace the master
curve precisely. The red squares are the experimental $q$ values quoted in
Fig.~4, placed on the curve at the phase $\Phi=2\,\mathrm{arccot}(q)$
implied by Eq.~(\ref{eq:master}) within their respective $2\pi$ window;
this placement illustrates the consistency of the measured
$(q,\Phi)$ assignments with the master curve (it is a consistency check
rather than an independent test, since the experimental $\Phi$ values are
themselves inferred from the measured line shapes). The experimental
observations of the main text continuous decrease of $q$ within a region, and reversal of the line-shape asymmetry are all features of Eq.~(\ref{eq:master}).

\section{Gate-voltage dispersion maps}
\label{sec:maps}

\begin{figure}[tb]
\includegraphics[width=0.95\textwidth]{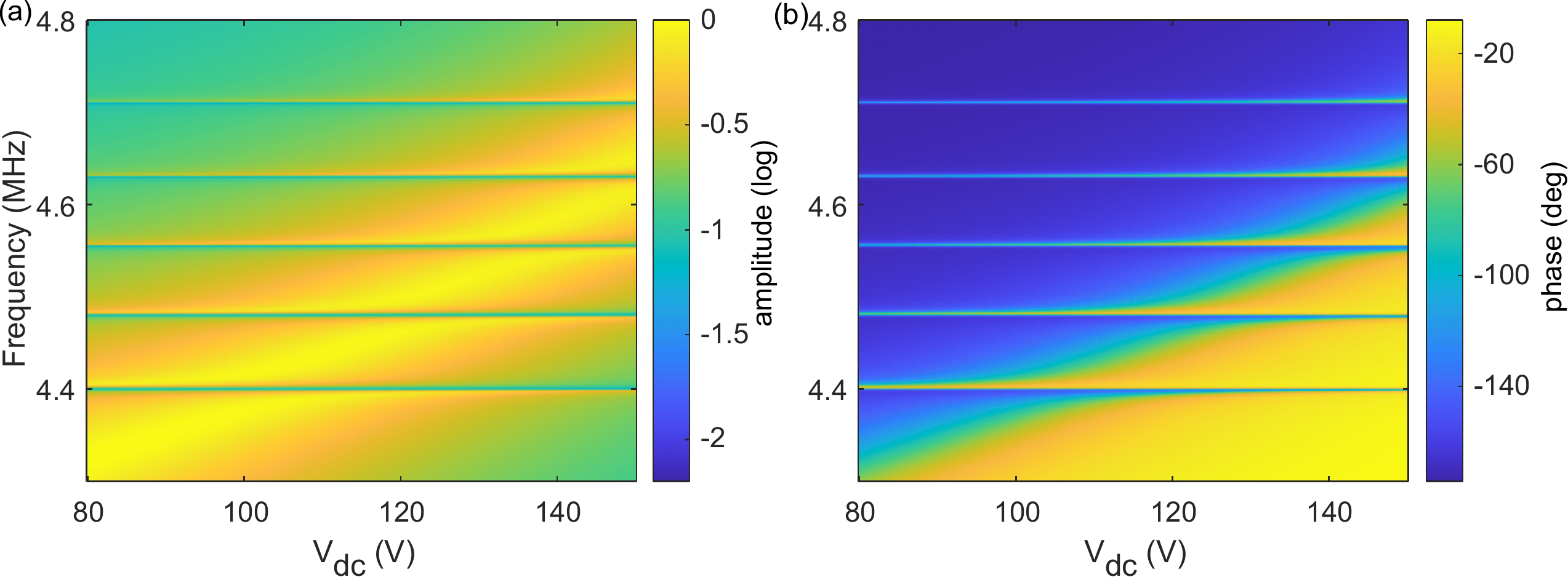}
\caption{Simulated gate-voltage dispersion maps for a graphene mode tuned via
Eq.~(\ref{eq:tuning}) across a comb of five fixed SiN$_x$ modes:
(a)~amplitude (log scale) and (b)~phase of $X_g$. Compare Fig.~2(A) of the main text: the bright graphene branch disperses with voltage
while dark antiresonance lines remain pinned at the fixed SiN$_x$
frequencies, with a phase disturbance at every crossing.}
\label{fig:S4}
\end{figure}

To connect with Fig.~2(A) of the main text, we compute the full
two-dimensional response $|X_g(\omega,V_{dc})|$ and its phase for a graphene
mode tuned according to
\begin{equation}
f_g(V_{dc})=\sqrt{f_0^{2}+\eta V_{dc}^{2}}\,,
\label{eq:tuning}
\end{equation}
the standard capacitive-tensioning form for a membrane in the
tension-stiffening regime [cf.\ the first term of Eq.~(2) of the main text],
with $(f_0,\eta)$ anchored to the end points of the measured dispersion in
Fig.~2(A) (85~V $\to$ 4.35~MHz, 150~V $\to$ 4.63~MHz), coupled to a comb of
five fixed SiN$_x$ modes. The result, Fig.~\ref{fig:S4}, reproduces the
experimental morphology: a bright graphene branch dispersing upward with
voltage, crossed by dark horizontal antiresonance lines pinned at the SiN$_x$
frequencies [cf.\ statement~(i) of Sec.~\ref{sec:model}: the zeros are
locked to $\omega_n'$ regardless of gate voltage], and a phase map showing a
localized phase disturbance at every crossing.

\section{Sensing metrics}
\label{sec:sensing}

\begin{figure}[tb]
\includegraphics[width=\textwidth]{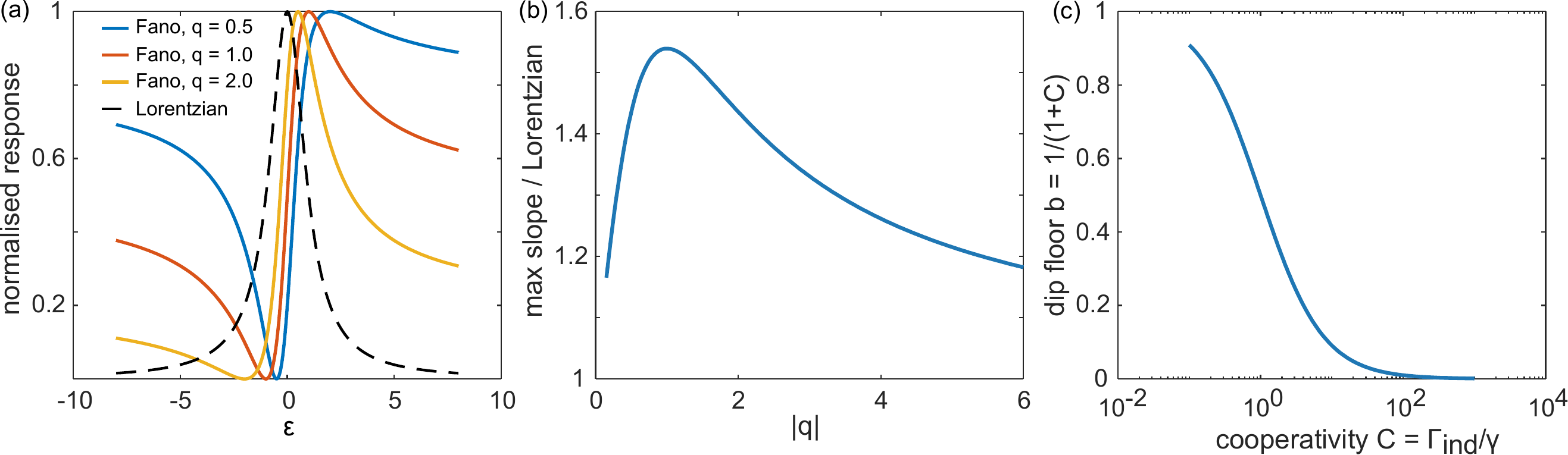}
\caption{Sensing metrics of the Fano line, Eq.~(\ref{eq:fanoprofile}) with
$b=0$ in (a), (b). (a)~Normalized Fano profiles for several $q$ compared with
a Lorentzian of equal linewidth. (b)~Maximum slope of the normalized profile
relative to the Lorentzian, versus $|q|$. (c)~Antiresonance dip floor
$b=1/(1+C)$ versus cooperativity, Eq.~(\ref{eq:qresult}).}
\label{fig:S5}
\end{figure}

Three quantitative consequences of Sec.~\ref{sec:fano} underpin the sensing
claims of the main text (Fig.~\ref{fig:S5}):

\begin{enumerate}
\item \textbf{Slope enhancement.} The steepest slope of the normalized Fano
profile exceeds that of a Lorentzian of the same total linewidth by a factor
that grows with $|q|$ [Fig.~\ref{fig:S5}(b)]; near $|q|\simeq1$--$3$, typical
of the measured features, the transduction slope (signal change per unit
frequency shift) is enhanced several-fold, directly improving the
responsivity of frequency-shift sensing.
\item \textbf{Pinned antiresonance.} The dip frequency is locked to the
SiN$_x$ mode [Sec.~\ref{sec:model}, statement~(i)], so tracking the dip
inherits the superb frequency stability of the high-$Q$ SiN$_x$ mode while
being interrogated through the strongly transduced, electrically tunable
graphene channel.
\item \textbf{Contrast as a cooperativity meter.} The dip floor
$b=1/(1+C)$ [Fig.~\ref{fig:S5}(c)] converts a simple amplitude reading into a
measurement of the coupling cooperativity $C$; conversely, any perturbation
of $\gamma_s$, $\gamma_g$, or the coupling (e.g., by adsorbed mass or stress)
registers as a change in dip depth.
\end{enumerate}

\section{Frequency locking of the driven Duffing graphene mode}
\label{sec:locking}

We now extend the model to the strong-drive regime of Fig.~5 of the main
text. Membrane resonators of atomic thickness stiffen under large
deflection, so at strong drive the graphene equation of motion acquires a
cubic restoring term $\beta x_g^{3}$ (hardening Duffing nonlinearity),
while the massive, high-tension SiN$_x$ modes remain linear at these
amplitudes. Writing $x_g(t)=\mathrm{Re}[A\,e^{i\omega t}]$ and keeping the
resonant (rotating-frame) term of the cubic, the complex amplitude obeys
\begin{equation}
\left[\,d_g(\omega)-\Sigma(\omega)+c_3|A|^{2}\,\right]A=\frac{F_0}{m_g},
\qquad
c_3=\frac{3\beta}{4m_g},
\label{eq:duffamp}
\end{equation}
with $d_g$ and $\Sigma$ exactly as in Eq.~(\ref{eq:exact}). Separating
real and imaginary parts, $x\equiv|A|^{2}$ solves the real cubic
\begin{equation}
c_3^{2}\,x^{3}+2Rc_3\,x^{2}+\left(R^{2}+I^{2}\right)x=P,
\qquad
R=\mathrm{Re}\!\left[d_g-\Sigma\right],\quad
I=\mathrm{Im}\!\left[d_g-\Sigma\right],\quad
P=\left(\frac{F_0}{m_g}\right)^{2},
\label{eq:duffcubic}
\end{equation}
which admits up to three positive solutions: the familiar Duffing
bistability, with the smallest and largest roots stable and the middle
root unstable. In a quasi-static upward frequency sweep the system remains
on the branch continuously connected to the previous frequency until that
branch ceases to exist at the saddle-node bifurcation, where the amplitude
collapses: this defines the jump-down frequency $\omega_J/2\pi$. For an isolated
Duffing resonator ($\Sigma=0$), $\omega_J/2\pi$ follows the backbone curve and
advances {continuously} with the drive.

The comb of discrete modes changes this qualitatively through the same
antiresonances that produce the Fano zeros in the linear regime. At the
$n$-th discrete frequency, $d_n(\omega_n')=i\gamma_n\omega_n'$, so the
self-energy contribution is purely imaginary,
$-\Sigma(\omega_n')=+\,i\lambda_n^{2}/(\gamma_n\omega_n')$, and it adds to
the {dissipative} part of the response:
\begin{equation}
I(\omega_n')=\gamma_g\omega_n'+\frac{\lambda_n^{2}}{\gamma_n\omega_n'}
=\gamma_g\omega_n'\left(1+C_n'\right),
\qquad
C_n'\equiv\frac{\lambda_n^{2}}{\gamma_g\gamma_n\omega_n'^{2}}
=\frac{C_n}{\sin^{2}\phi_n}\;,
\label{eq:barrier}
\end{equation}
where $C_n$ and $\phi_n$ are the cooperativity and continuum phase of
Secs.~\ref{sec:fano}--\ref{sec:modes}. Each SiN$_x$ mode therefore acts on
the nonlinear graphene branch as a narrow {damping barrier} of height
$(1+C_n')$. For the Duffing cubic~(\ref{eq:duffcubic}) the upper branch
exists at a given frequency only if the drive power exceeds a critical
value with the standard scaling $P_c\propto I^{3}/c_3$; consequently, when
the advancing jump-down point reaches a discrete mode, the upper branch is
destroyed {at} the antiresonance and $\omega_J/2\pi$ is pinned to
$\omega_n'/2\pi$. The pinning persists a locking plateau until the
drive exceeds the escape threshold
\begin{equation}
F_{\rm esc}\sim F_c\left(1+C_n'\right)^{3/2},
\label{eq:escape}
\end{equation}
at which point the branch survives the barrier and $\omega_J/2\pi$ snaps forward,
in practice directly to the next mode of the comb. The plateau length in
drive is therefore set by the cooperativity of the individual mode, which
explains the markedly unequal plateau widths of Fig.~5(B) of the main
text and their identification as a map of the mode-resolved coupling
strengths.

\begin{figure}[tb]
\includegraphics[width=0.9\textwidth]{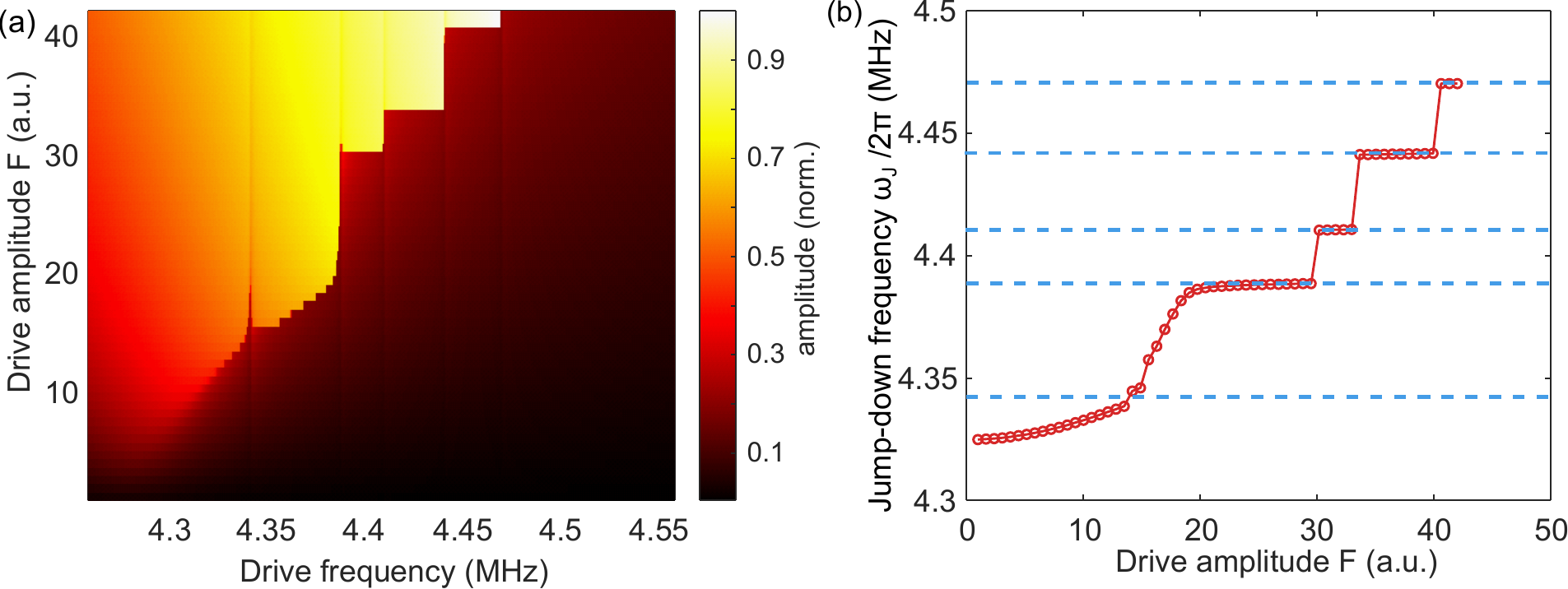}
\caption{Numerical demonstration of frequency locking of the Duffing
graphene mode to the SiN$_x$ comb
[Eqs.~(\ref{eq:duffamp})--(\ref{eq:escape})], to be compared with
Figs.~5(A) and 5(B) of the main text. (a)~Quasi-static sweep-up amplitude
versus drive frequency and drive amplitude $F$ for a graphene mode at
$\omega_g/2\pi=4.28$~MHz ($Q_g=80$) with hardening nonlinearity, coupled to five
discrete modes at the experimental plateau frequencies (dashed lines in
(b)). (b)~Extracted jump-down frequency $\omega_J/2\pi$ versus drive amplitude: the staircase locks onto the discrete-mode frequencies, with plateau lengths
set by the mode cooperativities $C_n'$.}
\label{fig:S6}
\end{figure}

Fig~\ref{fig:S6} shows a numerical demonstration with parameters
representative of the $V_{dc}=113$~V operating point ($\omega_g/2\pi=4.28$~MHz,
$Q_g=80$, five discrete modes at the plateau frequencies of Fig.~5(B) with
couplings $\lambda_n^{2}$ of order one in the units of
Sec.~\ref{sec:validation}). Panel~(a) maps the sweep-up amplitude versus
drive frequency and drive amplitude: the response reproduces the
morphology of Fig.~5(A) antiresonance lines pinned at the SiN$_x$
frequencies at all drive levels, and a bistability edge that advances in
discrete steps. Panel~(b) shows the extracted jump-down frequency: a
staircase whose plateaus sit on the discrete-mode frequencies. The weakly
coupled first mode, which the sweep reaches close to the onset of
bistability, produces only a brief pinning inflection, while the more
strongly coupled modes produce progressively longer plateaus, in
qualitative agreement with the experiment.

The seeding experiment of Fig.~5(C) admits a natural interpretation in
this picture. With the primary drive parked on a plateau, the system sits
just below the escape threshold~(\ref{eq:escape}) of the pinning mode. A
weak second tone applied at the locking point rings up additional
amplitude at the barrier and supplies the increment needed to carry the
upper branch across it; the switching is therefore threshold-like, the
system advances by exactly one plateau per threshold, and the thresholds
move to lower seeding amplitude as the primary drive is increased all
three features observed in Fig.~5(C). A quantitative two-tone (rotating
frame with two frequencies) treatment is beyond the scope of this
Supplementary.

\end{document}